\documentclass[letterpaper,twoside,twocolumn,american,showpacs,prl,aps,superscriptaddress]{revtex4}
\usepackage[T1]{fontenc}
\usepackage{color}
\usepackage{textcomp}
\usepackage{amstext}
\usepackage{amssymb}
\usepackage{graphicx}

\makeatletter

\pdfpageheight\paperheight
\pdfpagewidth\paperwidth

\newcommand{\lyxmathsym}[1]{\ifmmode\begingroup\def\b@ld{bold}
  \text{\ifx\math@version\b@ld\bfseries\fi#1}\endgroup\else#1\fi}

\@ifundefined{textcolor}{}
{%
 \definecolor{BLACK}{gray}{0}
 \definecolor{WHITE}{gray}{1}
 \definecolor{RED}{rgb}{1,0,0}
 \definecolor{GREEN}{rgb}{0,1,0}
 \definecolor{BLUE}{rgb}{0,0,1}
 \definecolor{CYAN}{cmyk}{1,0,0,0}
 \definecolor{MAGENTA}{cmyk}{0,1,0,0}
 \definecolor{YELLOW}{cmyk}{0,0,1,0}
}

\makeatother

\usepackage{babel}
\begin{document}

\title{Triplet-singlet conversion by broadband optical pumping}

\author{R. Horchani }

\affiliation{Laboratoire Aimé Cotton, CNRS, Université Paris-Sud, Bâtiment 505,
91405 Orsay, France}

\author{H. Lignier}

\affiliation{Laboratoire Aimé Cotton, CNRS, Université Paris-Sud, Bâtiment 505,
91405 Orsay, France}

\author{N. Bouloufa-Maafa}

\affiliation{Laboratoire Aimé Cotton, CNRS, Université Paris-Sud, Bâtiment 505,
91405 Orsay, France}

\affiliation{Université Cergy-Pontoise, 95000 Cergy-Pontoise, France}

\author{A. Fioretti}

\affiliation{Laboratoire Aimé Cotton, CNRS, Université Paris-Sud, Bâtiment 505,
91405 Orsay, France}

\author{P. Pillet}

\affiliation{Laboratoire Aimé Cotton, CNRS, Université Paris-Sud, Bâtiment 505,
91405 Orsay, France}

\author{D. Comparat}

\affiliation{Laboratoire Aimé Cotton, CNRS, Université Paris-Sud, Bâtiment 505,
91405 Orsay, France}

\date{\today}
\begin{abstract}
We demonstrate the conversion of cold Cs$_{2}$ molecules initially
distributed over several vibrational levels of the lowest triplet
state $a^{3}\Sigma_{u}^{+}$ into the singlet ground state $X^{1}\Sigma_{g}^{+}$.
This conversion is realized by a broadband laser exciting the molecules
to a well-chosen state from which they may decay to the singlet state
throug\textcolor{black}{h two sequential single-photon emission steps:
Th}e first photon populates levels with mixed triplet-singlet character,
making possible a second spontaneous emission down to several vibrational
levels of the $X^{1}\Sigma_{g}^{+}$ states. By adding an optical
scheme for vibrational cooling, a substantial fraction of molecules
are transferred to the ground vibrational level of the singlet state.
The efficiency of the conversion process, with and without vibrational
cooling, is discussed at the end of the article. The presented conversion
is general in scope and could be extended to other molecules. 
\end{abstract}


\maketitle
The last decade has witnessed increasing experimental efforts to produce
large samples of ultracold molecules in a well-defined quantum state.
Such samples constitute a very interesting basis for a great variety
of studies ranging from controlled molecular dynamics \cite{krems2008}
and anisotropic long-range interactions \cite{2009PCCP_Ospelkaus_dipolargas}
to precision measurements \cite{2009RepProgPhys_Dulieu} and quantum
computing \cite{Review_Ye2009}. Internal-state manipulation of ultracold
alkali-metal molecules has been in the spotlight with the stimulated
raman adiabatic passage (STIRAP) technique, which allows transfer,
with near-unity efficiency, of weakly bound molecules produced by
magneto-association in the lowest triplet state to the absolute singlet
ground state \cite{2008PhRvL_Denschlag_STIRAP,2008Science_Ospelkaus_STIRAP}.
However, this technique is not suited to samples of molecules distributed
in several vibrational levels, such as those produced by the widespread
technique of \textcolor{black}{cold-atom photoassociation. }

Motivated by the goal of finding general methods to achieve this transfer,
we propose an optical scheme to move a whole vibrational distribution
from a specific electronic state to another one of different multiplicity
and parity. By combining this technique with our vibrational cooling
technique \cite{2008Sci...321..232V}, we are able to produce large
samples of ultracold molecules in the lowest vibrational level of
the ground state. This successful combination demonstrates the versatility
of the optical vibrational cooling. Our demonstration relies on photoassociated
ultracold Cs$_{2}$ molecules stabilized in a vibrational distribution
of the lowest triplet state $a^{3}\Sigma_{u}^{+}$($a$) that a suitable
laser converts into the ground singlet state $X^{1}\Sigma_{g}^{+}$
($X$). Such a scheme provides an alternative way to efficiently produce
ground state molecules with photoassociation that mostly works in
forming molecules in the lowest triplet state \cite{2006RvMP...78..483J}.
In the case of our experiment, it forms twice the number of molecules
in the absolute vibrational ground level than any other PA scheme
\cite{2009PhRvA..79b1402V}. 

\begin{figure*}
\begin{centering}
\includegraphics[width=0.33\textwidth]{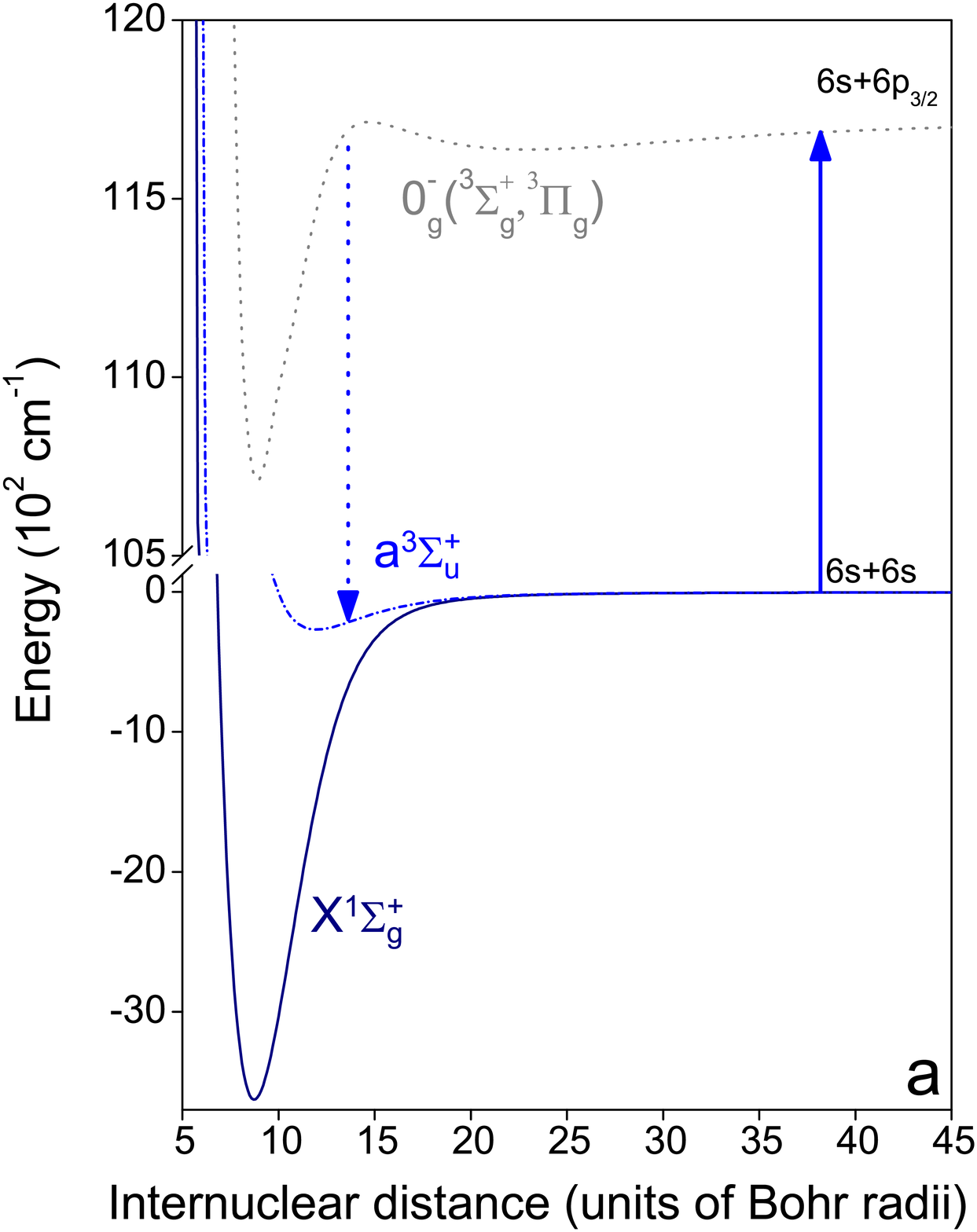}\includegraphics[width=0.33\textwidth]{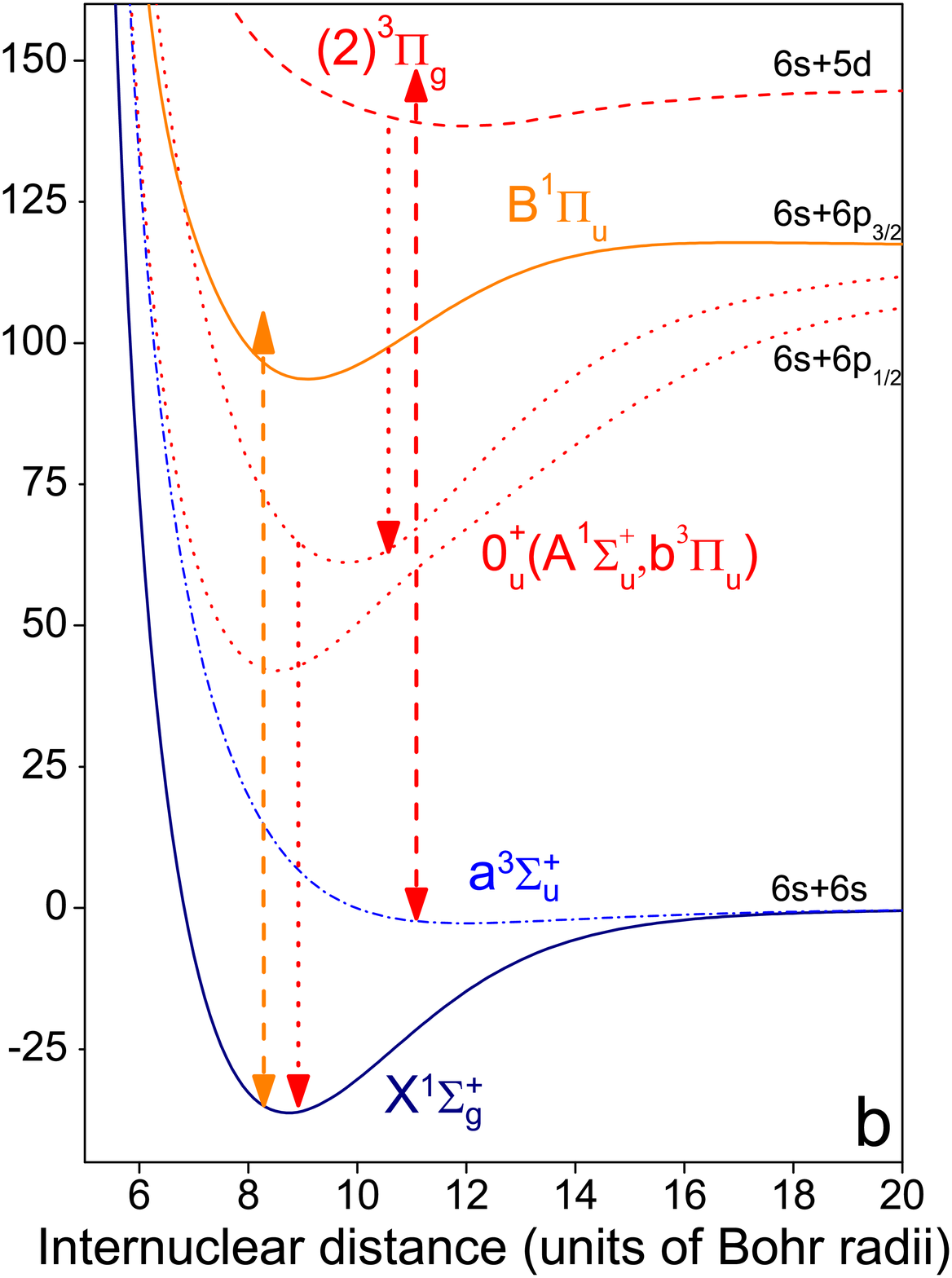}\includegraphics[width=0.33\textwidth]{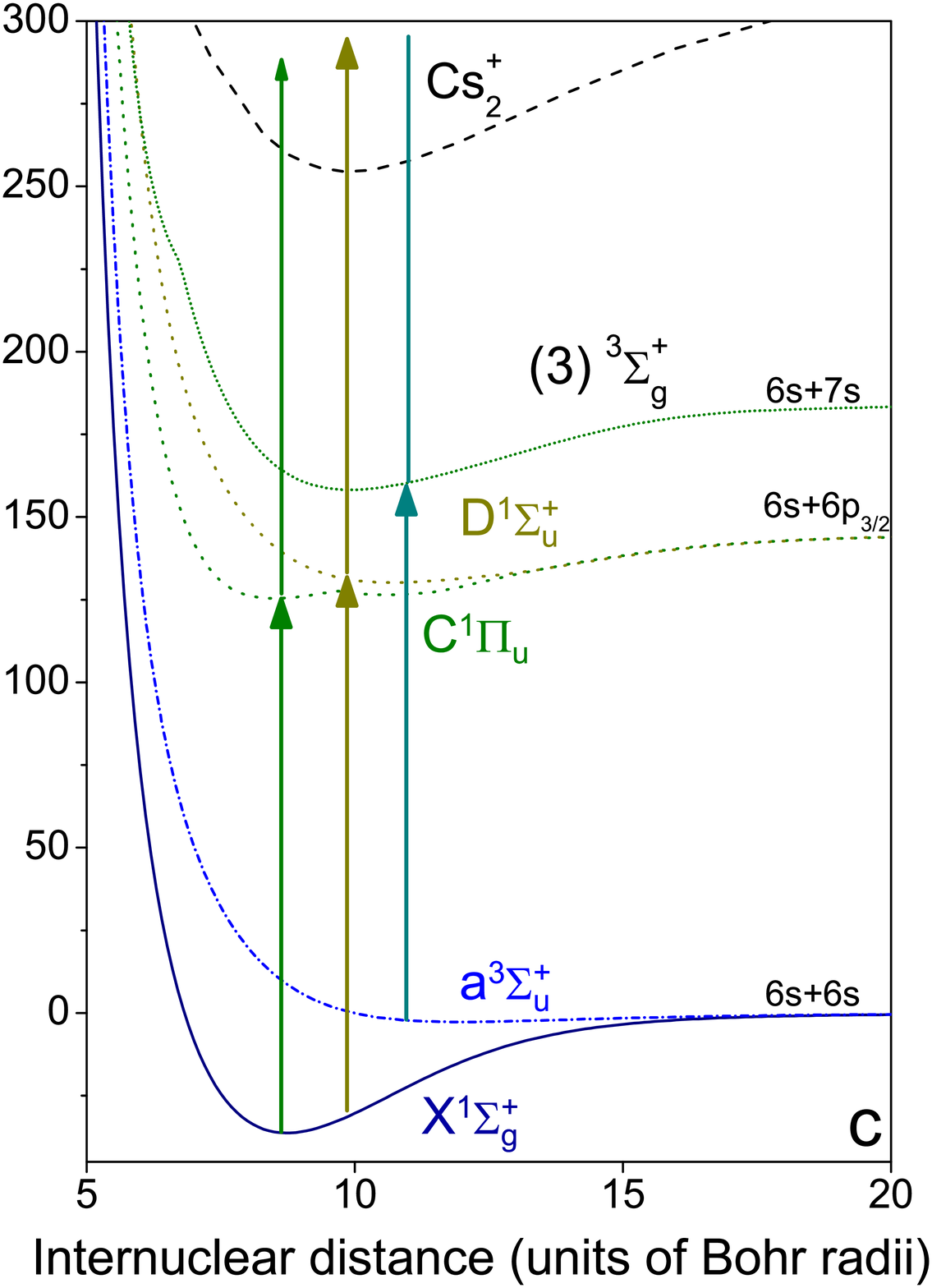}
\par\end{centering}

\caption{(Color online) Transitions used to a) photoassociate b) convert, cool
and c) detect Cs$_{2}$ molecules. a) Photoassociation: The PA laser
frequency (solid line up arrow) produces molecules in the $a$ state
after spontaneous emission (dotted down arrow). b) Conversion: the
laser used for conversion (dashed double arrow) transfers molecules
from the $a$ state to the $^{3}\Pi_{g}$($6s+5d$) state. Apart from
possible dissociation, two decay paths (dotted down arrows) may occur
via spontaneous emission: toward the $a$ state or toward the $X$
state by passing through the mixed $0_{u}^{+}$ states (two photons
are involved). c) Detection: either in the $a$ state or in the bottom
of the $X$ states, two-photon ionization allows detection of the
vibrational populations (solid up arrows). In the case of the $X$
state, the probed spectral range has two distinct possible transitions:
through either $C^{1}\Pi_{u}$ or $D^{1}\Sigma_{u}^{+}$.\label{fig:figure1}}

\end{figure*}

The triplet-singlet conversion is not straightforward because of the
quantum selection rules imposed by electric dipole transitions. The
multiplicity $2S+1$, where $S$ is the total spin quantum number,
must be conserved, and moreover, in the case of homonuclear molecules,
the \emph{ungerade-gerade} parity must be changed ($u\leftrightarrow g$).
The conversion of multiplicity can occur if an intermediate state
of the transition process is a mixture of the initial and final multiplicities
due to, for instance, a spin-orbit coupling. The choice of the transition
steps, based on the study published in \cite{2011_PRA_Bouloufa_conversion},
is summarized in Fig. \ref{fig:figure1}(b): Cs$_{2}$ molecules prepared
in the $a$ state are excited by a broadband laser to the $(2)^{3}\Pi_{g}$
state from which they may decay in two steps: first to the intermediate
mixed state $0_{u}^{+}(A^{1}\Sigma_{u}^{+}+b^{3}\Pi_{u})$ and then
to the $X$ state. This decay channel is not unique and molecules
may dissociate or simply go back to the $a$ state. It is important
to note that the fraction of molecules going back to the $a$ state
can be re-excited as long as the broadband laser contains the suitable
frequencies. In other words, the large laser bandwidth has two closely
related roles: exciting several levels of the initial vibrational
distribution and recycling decayed molecules in the $a$ state.

Our experimental setup is based on a classical magneto-optical trap
(MOT). It provides an atomic cloud of $\sim10^{7}$ atoms at a temperature
of $\sim150\,\mu$K with a peak density of $\sim10^{11}$ atoms$/$cm$^{3}$.
An experiment cycle lasts $100$ ms. During the first $50$ ms, three
different lasers, simultaneously switched on and used for PA, internal
conversion and vibrational cooling. As molecules are not trapped,
they fall freely under the influence of gravity and are available
for manipulation and detection during a period of about $10$ ms.
We now describe these important steps in more detail. 

The detection stage is realized by resonance enhanced 2-photon ionization
(RE2PI). To this end, we use a pulsed dye (DCM) laser pumped by the
second harmonic of a pulsed Nd:YAG laser with a $7$-ns pulse duration
an\textcolor{black}{d $0.5$-cm$^{\lyxmathsym{\textminus}1}$ l}inewidth
($2$ mJ per pulse). The two-stage ionization produces Cs$_{2}^{+}$
ions that are detected by three stacked microchannel plates. The spectroscopic
signal is monitored on a fast oscilloscope performing an averaging
over 10 cycles. By scanning the RE2PI wave number in the available
range ($15800-16100$ cm$^{-1}$), we obtain an ionization spectrum
whose lines reveal vibrational transitions either from the $a$ state
via the $(3)^{3}\Sigma_{g}^{+}$ state \cite{2010MolPh.108.2355B},
or the $X$ state via the $C^{1}\Pi_{u}$ or $D^{1}\Sigma_{u}^{+}$
state {[}Fig. \ref{fig:figure1}(c){]} \cite{1988_JCP_demtroder_Cs2_sepctroscopy}.

The photoassociation is achieved by focusing a $50$-ms, $750$-mW
Ti:sapphire laser onto the MOT. When the PA wavelength is tuned into
resonance, pairs of colliding atoms absorb a photon to form molecules
in an excited state, which are then stabilized by spontaneous emission
{[}Fig. \ref{fig:figure1}(a){]}. In the following we consider only
PA schemes which mainly populate the $a$ state. We mainly used an
efficient PA scheme identified by Fioretti \emph{et al. }\cite{1999EPJD....5..389F},
denoted $G_{1}$, at the PA wave number $\tilde{\nu}=11730.0422$
cm$^{-1}$. How the vibrational levels of the $a$ state are populated
depends upon the Franck-Condon (FC) factors between its $54$ vibrational
levels \cite{2009_JCP_LiLi_triplet} and the PA excited level. To
know the experimental vibrational distribution resulting from a given
PA scheme, we performed RE2PI spectroscopy.

\begin{figure}
\begin{centering}
\includegraphics[width=1\columnwidth]{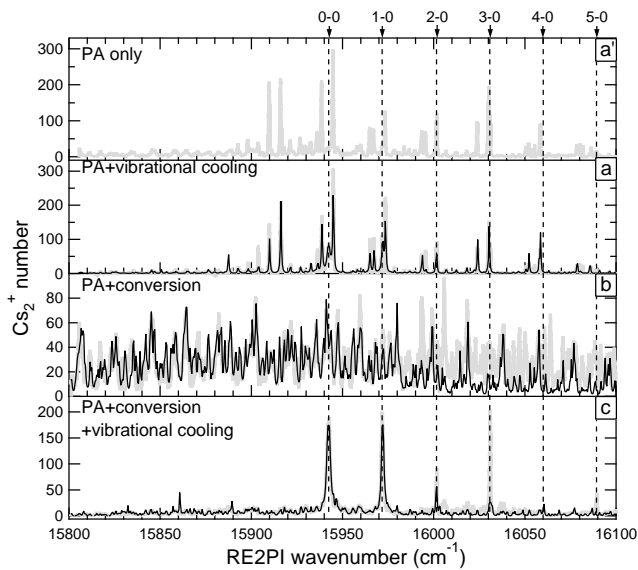}
\par\end{centering}

\caption{\textcolor{black}{RE2PI spectra (thick gray lines) and their fits
(thin black lines) in four cases: (a$^{\prime}$) PA spectrum revealing
the lines from the $a^{3}\Sigma_{u}^{+}$ state (fit not shown). (a)
PA with vibrational cooling: Thus, the vibrational series of the $X$
state has been reduced to a few lines corresponding to the $v_{X}=0$
population. (b) When the conversion laser is added to PA only, the
previous spectrum undergoes strong modifications. (c) The same as
(b) with vibrational pumping: Intense peaks corresponding to $v_{X}=0$
emerge whereas the lines from the $a^{3}\Sigma_{u}^{+}$ state have
apparently disappeared. Above the graph, the numbers associated to
the arrows represent the vibrational numbers of the $(C,v_{C})\leftarrow(X,v_{X}=0)$
transitions.}\textcolor{red}{{} \label{fig:Spectra}}}
\end{figure}
\begin{figure}
\begin{centering}
\includegraphics[width=0.8\columnwidth]{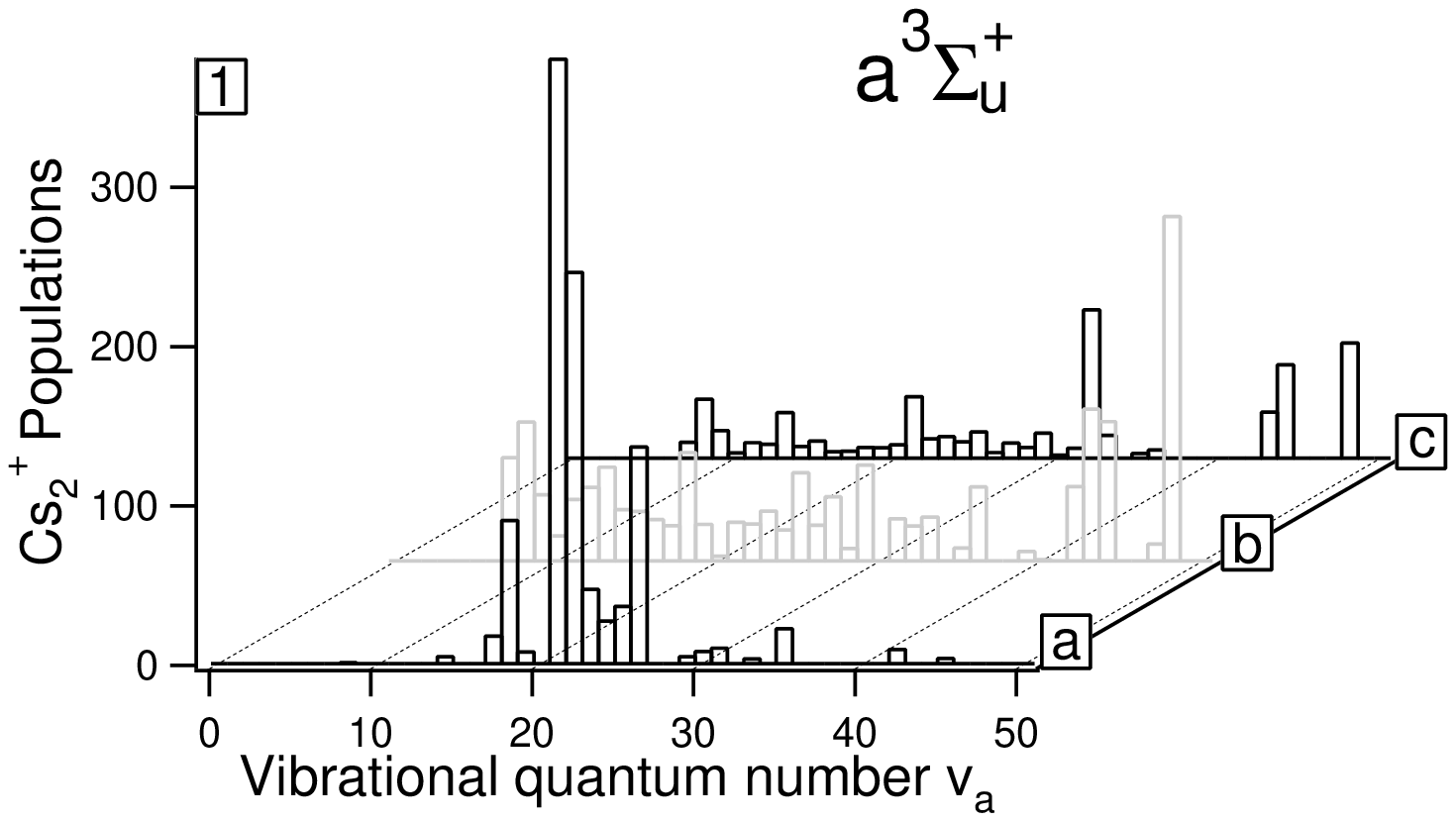}
\par\end{centering}

\begin{centering}
\includegraphics[width=0.8\columnwidth]{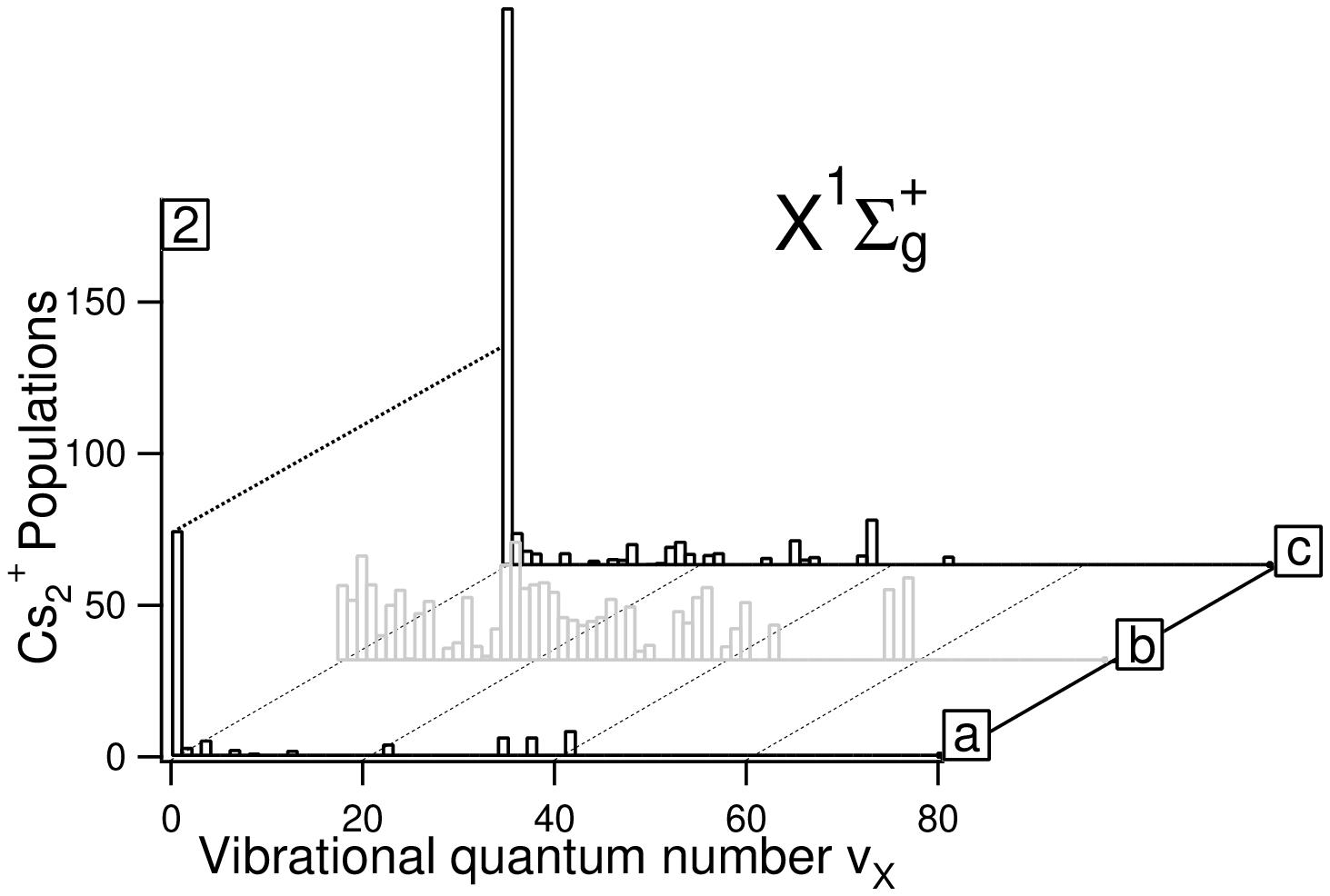}
\par\end{centering}

\caption{Vibrational distributions of the $a$ state (1) and the $X$ state
(2) obtained by fitting the spectra of Figs. \ref{fig:Spectra}(a),
(b) and (c). The distributions (a) and (c) are well determined given
that the line densities of their respective spectra are low enough
to enable a reliable fit. Here the use of vibrational cooling shows
that the population of $v_{X}=0$ roughly doubles when the conversion
laser is applied. The quality of the (b) distribution is questionable
as the integrated number of molecules is distinctly too high.\label{fig:distributions_a_X}}
\end{figure}

The triplet-singlet conversion is performed by femtosecond laser pulses
($500$ mW, $120$ fs) tuned to $\sim13910$ cm$^{-1}$ in order to
induce $\left[(2)^{3}\Pi_{g},v_{\Pi})\right]\leftarrow(a,v_{a})$
transitions. As mentioned above, thanks to a two-photon cascade, the
excited molecules are likely to end in the $X$ state {[}Fig. \ref{fig:figure1}(b){]},
thus realizing the intended conversion. Here the spectral width \textcolor{black}{($\sim200$
cm$^{-1}$)}\textcolor{red}{{} }is the only interesting feature: It
is larger than the energy range of the vibrational distribution ($\sim50$
cm$^{-1}$), which implies that all the vibrational populations in
the $a$ state are likely to be excited. This laser does not affect
the populations in the $X$ state.

A vibrational cooling scheme is then employed to pump converted molecules
into $v_{X}=0$ \cite{2008Sci...321..232V}. For this purpose, we
use a laser diode running below its lasing threshold characterized
by a spectral width of about $200$ cm$^{-1}$. This light, with a
spectrum centered at $12940$ cm$^{-1}$, causes multiple transitions
between the $X$ state and the $B^{1}\Pi_{u}$ state {[}Fig. \ref{fig:figure1}(b){]}.
Because the part of the spectrum inducing transitions from $v_{X}=0$
is removed by an interferential filter, molecules are progressively
pumped down to that level \cite{2009PRA_Sofikitis}.\textcolor{black}{{}
As expected intuitively, as the laser spectral width becomes broader,
the vibrational levels are pumped more efficiently. According to numerical
simulations, }a $200$ cm$^{-1}$\textcolor{black}{{} linewidth enables
one to transfer a population roughly spread over the $10$ first vibrational
levels to $v_{X}=0$ with an efficiency greater than $90$ \%. }Molecules
lying in higher levels also undergo transitions, but they slowly spread
throughout the vibrational levels of the $X$ state. We checked that
vibrational cooling does not disturb the formation and conversion
of molecules in the $a$ state, meaning that all these processes can
work simultaneously.

\textcolor{black}{In order to demonstrate molecular conversion and
vibrational cooling, four relevant configurations are discussed: (a$^{\prime}$)
PA only, (a) PA and vibrational cooling , (b) PA and laser conversion,
and (c) PA, conversion laser, and vibrational cooling. The corresponding
RE2PI spectra have been systematically recorded and subjected to a
fitting procedure considering the possible vibrational transitions
shown in Fig. \ref{fig:figure1}(c) and their characteristics \cite{2010MolPh.108.2355B,1988_JCP_demtroder_Cs2_sepctroscopy},
that is, the transition energies, the FC factors, the average electrical
transition dipole moments (TDM), and also the linewidth of the pulsed
dye laser. This gives us the vibrational populations of the $a$ and
$X$ states for each configuration.}

In the (a) configuration, we obtained the typical RE2PI spectrum shown
in Fig. \ref{fig:Spectra}(a). Although the PA scheme essentially
produces molecules in the $a$ state, a fraction of those are stabilized
in the $X$ state through a two-photon decay process (equivalent to
\cite{2011PCCP_Lignier,2008Sci...321..232V}). In order to quantify
the number of molecules initially in the $X$ state, we added the
vibrational cooling to PA and obtained the spectrum shown in Fig.
\ref{fig:Spectra}(b). Although the RE2PI spectra obtained with and
without vibrational cooling look very similar, we note that a few
peaks appear when vibrational cooling is used. The positions of these
peaks exactly correspond to $(C,v_{C})\leftarrow(X,v_{X}=0)$ transitions,
which confirms that the vibrational populations of the $X$ state
are pumped to $v_{X}=0$. The rest of the spectrum (i.e., the lines
belonging to the vibrational series of the $a$ state) is not affected
by this manipulation. The interest of such a spectral ``cleaning''
is to improve the results of our fitting procedure and make molecules
visible that will not be mistaken for those that will result from
the conversion process.  The fitting procedure indicates, with good
reliability, that $90$ \% of molecules decayed in the $a$ state
are gathered in the range $18\lesssim v_{a}\lesssim26$ {[}Fig. \ref{fig:distributions_a_X}(a){]},
while the rest can be put into the $v_{X}=0$ level {[}Fig \ref{fig:distributions_a_X}.(a){]}.

In the (b) configuration, the RE2PI spectrum is deeply modified as
it is visible in the spectrum shown in Fig. \ref{fig:Spectra}(b).
This spectrum shows a high line density suggesting that the initial
populations of the $a$ state have been redistributed among numerous
levels. A visual inspection of the spectrum ensures that a part of
the molecules reaches the $X$ state because many lines are found
below $15827$ cm$^{-1}$ which is the cutoff wave number under which
there is no transition from the $a$ state. Due to the partial knowledge
of the transition characteristics, our fitting procedure encounters
limitations, especially for spectral regions above $16000$ cm$^{-1}$.\textcolor{black}{{}
Yet it shows that the $X$ state is much more populated than previously
as suggested by the comparison between the distributions (a) and (b)
in Fig. \ref{fig:distributions_a_X}(b).}

The (c) configuration, where vibrational cooling is added to the conversion,
definitively demonstrates that molecules have been transferred to
the $X$ state. We effectively note that the RE2PI spectrum, displayed
in Fig. \ref{fig:Spectra}(c), consists of intense peaks, proving
an accumulation of molecules in $v_{X}=0$, and a background of small
lines arising from molecules that escape from either the vibrational
pumping or the conversion scheme. The results of the fitting procedure
shown in Fig. \ref{fig:distributions_a_X}(c), compared to the situation
of Fig. \ref{fig:distributions_a_X}(a), confirm that the population
in $v_{X}=0$ has clearly increased while the distribution in the
$a$ state has decreased and changed.

\textcolor{black}{With this, it is clear that molecules undergo a
triplet-singlet conversion under the action of our scheme. The fitting
procedure indicates that the increment of $v_{X}=0$ population compared
to the initial number of molecules in the $a$ state reaches about
$10$ \% when the laser conversion and vibrational cooling process
are applied. We can deduce that $10$\% is a lower limit to the efficiency
of the sole conversion toward the $X$ state since the vibrational
cooling is not $100\%$ efficient. Our fitting procedure might be
considered as a good and general method to estimate this efficiency.
However, when the spectrum is too dense, like that obtained with configuration
(b), the vibrational distributions obtained are not reliable.}

\begin{figure}
\begin{centering}
\includegraphics[width=0.8\columnwidth]{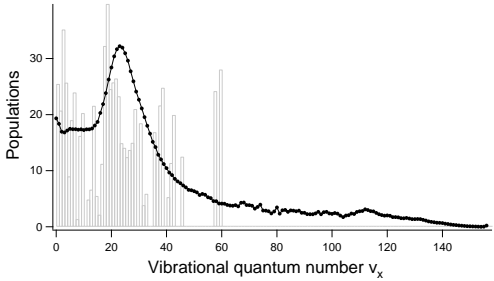}
\par\end{centering}

\caption{Comparison of the vibrational distributions of the $X$ state obtained
by the numerical simulation (dots) and the fitting procedure (bars,
same as Fig. \ref{fig:distributions_a_X}.2(b)) when only the conversion
process is applied. The absolute numbers of molecules per $v_{X}$
are different but the general behaviors are similar: The low levels
are mostly populated whereas there are few molecules above $v_{X}\approx40$.\label{fig:Comparison_fit_simulation}}
\end{figure}

\textcolor{black}{To improve our understanding and confirm our raw
observations, we numerically simulated the conversion process and
vibrational cooling. This simulation starts from the initial vibrational
distribution of the $a$ state, displayed in Fig. \ref{fig:distributions_a_X}(a),
and computes the evolution of the initial population submitted to
the conversion laser. According to its wavelength and width, this
laser enables to reach only the $0_{g}^{\pm}$ components; that is,
the lowest fine structure components of the $(2)^{3}\Pi_{g}$ state
\cite{Kraemer1997391}. The spontaneous emission from the $0_{g}^{\pm}$
components can then populate the $0_{u}^{+}(A^{1}\Sigma_{u}^{+},b^{3}\Pi_{u})$,
$0_{u}^{-}((2){}^{3}\Sigma_{u}^{+},b^{3}\Pi_{u})$ and $1_{u}((2){}^{3}\Sigma_{u}^{+},b^{3}\Pi_{u},B^{1}\Pi_{u})$,
all of them being components of the$(b)^{3}\Pi_{u}$ state. We consider
that $0_{u}^{-}$ and $1_{u}$ states are both metastable: The only
possible decay respectively occurs by quadrupolar transition (to the
$a$ state) and weak dipolar transition (to the $X$ state) due to
the $B^{1}\Pi_{u}$ component. We assume that the lifetime of these
states is so long that molecules reaching them are lost in the experiment.
This assumption is supported by the fact that we do not see any long
time dynamics in the conversion process. On the contrary, the $0_{u}^{+}$
state can efficiently decay to the $X$ state through its $A^{1}\Sigma_{u}^{+}$
component. Consequently, our simulation considers the only path leading
to the $X$ state in times that make the experimental detection possible:
$(2)^{3}\Pi_{g}\longleftrightarrow a$, $(2)^{3}\Pi_{g}\rightarrow0_{u}^{+}(A^{1}\Sigma_{u}^{+},b^{3}\Pi_{u})$,
$0_{u}^{+}(A^{1}\Sigma_{u}^{+},b^{3}\Pi_{u})\rightarrow X$. We took
into account the width of the conversion laser, the transition energies,
the FC factors and the average TDMs, ingredients that can be found
in more detail in \cite{2011_PRA_Bouloufa_conversion}. We note that
molecules excited in the $(2)^{3}\Pi_{g}$ state preferentially decay
back to the $a$ state because the TDM of this transition is about
nine times larger than the TDM of the transition leading to $0_{u}^{+}(A^{1}\Sigma_{u}^{+},b^{3}\Pi_{u})$.
This unfavorable feature does not really affect the whole conversion
process because the width of the conversion laser enables molecular
repumping from almost all the vibrational levels of the $a$ state
until they decay to $0_{u}^{+}$. Only the last vibrational levels
of the $a$ state near the dissociation limit have extremely weak
FC factors and are not recycled. Yet, as the distribution of the FC
factors does not favor the accumulation of molecules in these high
levels, only $3$\% of the molecules are lost in this way. Our simulation
also takes into account the dissociation events that occur in small
proportions ($10$\%). The remaining molecules ($87$\%) decay to
the intermediate states and a fraction of them eventually end up in
the $X$ state. In Fig. \ref{fig:Comparison_fit_simulation}, the
vibrational distribution in the $X$ state obtained by the fitting
procedure {[}Fig. \ref{fig:distributions_a_X}(b){]} and the numerical
simulation are compared. Although the populations for a given $v_{X}$
are different, both are spread over the same region of the $X$ state
(i.e., between $v_{X}=0$ and $v_{X}\approx40$). These differences
between the populations are not surprising due to the limitations
of the fitting procedure and the use of imperfect parameters in the
numerical simulation. Finally, by adding the vibrational cooling to
our simulation, we find that about $40$\% of the molecules reaching
the $X$ state are accumulated in $v_{X}=0$. A marginal part is lost
by dissociation and the rest by diffusion throughout the high vibrational
levels of the $X$ state. This implies that a higher value could be
reached by using a broader linewidth laser for the vibrational cooling
scheme. However we find that the number of molecules reaching $v_{X}=0$
is about three times greater in the simulation than in the experiment.
This expected discrepancy supports the idea that some molecules are
blocked in the metastable states and remain undetected. }

\textcolor{black}{We also experimentally tested the robustness of
the cooling and conversion process with other initial vibrational
distributions in the $a$ state by using other PA schemes. We used
transitions toward $\left[0_{g}^{-}(P_{3/2}),\, v=6,\, J=2\right]$
at $\tilde{\nu}=11665.2055$ cm$^{-1}$\cite{1999EPJD....5..389F}
and $\left[0_{g}^{-}(P_{1/2}),\, v=32\right]$ at $\tilde{\nu}=11158.5252$
cm$^{-1}$ \cite{2011PCCP_Lignier} that respectively form molecules
in $37\lesssim v_{a}\lesssim50$ and in the highest levels of the
$a$ state. The latter distribution is not transferred by our conversion
scheme, which is consistent with the previously evoked poor FC factors
between the $a$ state and $(2)^{3}\Pi_{g}$. On the other hand, the
distribution with $37\lesssim v_{a}\lesssim50$ is converted in similar
proportion to $G_{1}$. }

\textcolor{black}{In this work, we have been able to produce a large
sample of molecules in the absolute vibrational ground state, albeit
formed by PA in the lowest triplet state. The conversion process relies
on the proposal described in}\textcolor{black}{\emph{ }}\textcolor{black}{Bouloufa}\textcolor{black}{\emph{
et al. }}\textcolor{black}{\cite{2011_PRA_Bouloufa_conversion}, except
that, as suggested at the end of the same article, we make use of
a broadband rather than a single-mode conversion laser. This modification
explains why we detect $20$ times more molecules in the $X$ state
and, if vibrational cooling is added, even $700$ times more in $v_{X}=0$
than predicted in \cite{2011_PRA_Bouloufa_conversion}. On the other
hand, it must be noted that the conversion process and vibrational
cooling increase the rotational temperature. The reason is that each
photon absorption or emission is likely to change the rotational quantum
number by unity, and thus many transition processes provoke a population
spread over of the rotational levels (i.e., a rotational heating).}

We thank Olivier Dulieu for fruitful discussions, Myreille Aymar for
providing potential curves and TDMs and Maria Allegrini and Benjamin
Le Crom for their help with the experiment. Laboratoire Aimé Cotton
is a member of Institut Francilien de Recherche sur les Atomes Froids
(IFRAF) and of the LABEX PALM initiative. A. Fioretti has been supported
by the "Triangle de la Physique" under Contracts 2007-n.74T and
2009-035T "GULFSTREAM".


\begin{thebibliography}{17}
\expandafter\ifx\csname natexlab\endcsname\relax\def\natexlab#1{#1}\fi
\expandafter\ifx\csname bibnamefont\endcsname\relax
  \def\bibnamefont#1{#1}\fi
\expandafter\ifx\csname bibfnamefont\endcsname\relax
  \def\bibfnamefont#1{#1}\fi
\expandafter\ifx\csname citenamefont\endcsname\relax
  \def\citenamefont#1{#1}\fi
\expandafter\ifx\csname url\endcsname\relax
  \def\url#1{\texttt{#1}}\fi
\expandafter\ifx\csname urlprefix\endcsname\relax\def\urlprefix{URL }\fi
\providecommand{\bibinfo}[2]{#2}
\providecommand{\eprint}[2][]{\url{#2}}

\bibitem[{\citenamefont{Krems}(2008)}]{krems2008}
\bibinfo{author}{\bibfnamefont{R.~V.} \bibnamefont{Krems}},
  \bibinfo{journal}{Phys. Chem. Chem. Phys.} \textbf{\bibinfo{volume}{10}},
  \bibinfo{pages}{4079} (\bibinfo{year}{2008}).

\bibitem[{\citenamefont{Ni et~al.}(2009)\citenamefont{Ni, Ospelkaus, Nesbitt,
  Ye, and Jin}}]{2009PCCP_Ospelkaus_dipolargas}
\bibinfo{author}{\bibfnamefont{K.-K.} \bibnamefont{Ni}},
  \bibinfo{author}{\bibfnamefont{S.}~\bibnamefont{Ospelkaus}},
  \bibinfo{author}{\bibfnamefont{D.~J.} \bibnamefont{Nesbitt}},
  \bibinfo{author}{\bibfnamefont{J.}~\bibnamefont{Ye}}, \bibnamefont{and}
  \bibinfo{author}{\bibfnamefont{D.~S.} \bibnamefont{Jin}},
  \bibinfo{journal}{Phys. Chem. Chem. Phys.} \textbf{\bibinfo{volume}{11}},
  \bibinfo{pages}{9626} (\bibinfo{year}{2009}).

\bibitem[{\citenamefont{Dulieu and Gabbanini}(2009)}]{2009RepProgPhys_Dulieu}
\bibinfo{author}{\bibfnamefont{O.}~\bibnamefont{Dulieu}} \bibnamefont{and}
  \bibinfo{author}{\bibfnamefont{C.}~\bibnamefont{Gabbanini}},
  \bibinfo{journal}{Reports on Progress in Physics}
  \textbf{\bibinfo{volume}{72}}, \bibinfo{pages}{086401}
  (\bibinfo{year}{2009}).

\bibitem[{\citenamefont{Carr et~al.}(2009)\citenamefont{Carr, DeMille, Krems,
  and Ye}}]{Review_Ye2009}
\bibinfo{author}{\bibfnamefont{L.}~\bibnamefont{Carr}},
  \bibinfo{author}{\bibfnamefont{D.}~\bibnamefont{DeMille}},
  \bibinfo{author}{\bibfnamefont{R.}~\bibnamefont{Krems}}, \bibnamefont{and}
  \bibinfo{author}{\bibfnamefont{J.}~\bibnamefont{Ye}}, \bibinfo{journal}{New
  Journal of Physics} \textbf{\bibinfo{volume}{11}}, \bibinfo{pages}{055049}
  (\bibinfo{year}{2009}).

\bibitem[{\citenamefont{{Lang} et~al.}(2008)\citenamefont{{Lang}, {Winkler},
  {Strauss}, {Grimm}, and {Denschlag}}}]{2008PhRvL_Denschlag_STIRAP}
\bibinfo{author}{\bibfnamefont{F.}~\bibnamefont{{Lang}}},
  \bibinfo{author}{\bibfnamefont{K.}~\bibnamefont{{Winkler}}},
  \bibinfo{author}{\bibfnamefont{C.}~\bibnamefont{{Strauss}}},
  \bibinfo{author}{\bibfnamefont{R.}~\bibnamefont{{Grimm}}}, \bibnamefont{and}
  \bibinfo{author}{\bibfnamefont{J.~H.} \bibnamefont{{Denschlag}}},
  \bibinfo{journal}{Physical Review Letters} \textbf{\bibinfo{volume}{101}},
  \bibinfo{pages}{133005} (\bibinfo{year}{2008}).

\bibitem[{\citenamefont{{Ni} et~al.}(2008)\citenamefont{{Ni}, {Ospelkaus}, {de
  Miranda}, {Pe'er}, {Neyenhuis}, {Zirbel}, {Kotochigova}, {Julienne}, {Jin},
  and {Ye}}}]{2008Science_Ospelkaus_STIRAP}
\bibinfo{author}{\bibfnamefont{K.}~\bibnamefont{{Ni}}},
  \bibinfo{author}{\bibfnamefont{S.}~\bibnamefont{{Ospelkaus}}},
  \bibinfo{author}{\bibfnamefont{M.~H.~G.} \bibnamefont{{de Miranda}}},
  \bibinfo{author}{\bibfnamefont{A.}~\bibnamefont{{Pe'er}}},
  \bibinfo{author}{\bibfnamefont{B.}~\bibnamefont{{Neyenhuis}}},
  \bibinfo{author}{\bibfnamefont{J.~J.} \bibnamefont{{Zirbel}}},
  \bibinfo{author}{\bibfnamefont{S.}~\bibnamefont{{Kotochigova}}},
  \bibinfo{author}{\bibfnamefont{P.~S.} \bibnamefont{{Julienne}}},
  \bibinfo{author}{\bibfnamefont{D.~S.} \bibnamefont{{Jin}}}, \bibnamefont{and}
  \bibinfo{author}{\bibfnamefont{J.}~\bibnamefont{{Ye}}},
  \bibinfo{journal}{Science} \textbf{\bibinfo{volume}{322}},
  \bibinfo{pages}{231} (\bibinfo{year}{2008}).

\bibitem[{\citenamefont{{Viteau} et~al.}(2008)\citenamefont{{Viteau}, {Chotia},
  {Allegrini}, {Bouloufa}, {Dulieu}, {Comparat}, and
  {Pillet}}}]{2008Sci...321..232V}
\bibinfo{author}{\bibfnamefont{M.}~\bibnamefont{{Viteau}}},
  \bibinfo{author}{\bibfnamefont{A.}~\bibnamefont{{Chotia}}},
  \bibinfo{author}{\bibfnamefont{M.}~\bibnamefont{{Allegrini}}},
  \bibinfo{author}{\bibfnamefont{N.}~\bibnamefont{{Bouloufa}}},
  \bibinfo{author}{\bibfnamefont{O.}~\bibnamefont{{Dulieu}}},
  \bibinfo{author}{\bibfnamefont{D.}~\bibnamefont{{Comparat}}},
  \bibnamefont{and} \bibinfo{author}{\bibfnamefont{P.}~\bibnamefont{{Pillet}}},
  \bibinfo{journal}{Science} \textbf{\bibinfo{volume}{321}},
  \bibinfo{pages}{232} (\bibinfo{year}{2008}).

\bibitem[{\citenamefont{{Jones} et~al.}(2006)\citenamefont{{Jones}, {Tiesinga},
  {Lett}, and {Julienne}}}]{2006RvMP...78..483J}
\bibinfo{author}{\bibfnamefont{K.~M.} \bibnamefont{{Jones}}},
  \bibinfo{author}{\bibfnamefont{E.}~\bibnamefont{{Tiesinga}}},
  \bibinfo{author}{\bibfnamefont{P.~D.} \bibnamefont{{Lett}}},
  \bibnamefont{and} \bibinfo{author}{\bibfnamefont{P.~S.}
  \bibnamefont{{Julienne}}}, \bibinfo{journal}{Reviews of Modern Physics}
  \textbf{\bibinfo{volume}{78}}, \bibinfo{pages}{483} (\bibinfo{year}{2006}).

\bibitem[{\citenamefont{{Viteau} et~al.}(2009)\citenamefont{{Viteau}, {Chotia},
  {Allegrini}, {Bouloufa}, {Dulieu}, {Comparat}, and
  {Pillet}}}]{2009PhRvA..79b1402V}
\bibinfo{author}{\bibfnamefont{M.}~\bibnamefont{{Viteau}}},
  \bibinfo{author}{\bibfnamefont{A.}~\bibnamefont{{Chotia}}},
  \bibinfo{author}{\bibfnamefont{M.}~\bibnamefont{{Allegrini}}},
  \bibinfo{author}{\bibfnamefont{N.}~\bibnamefont{{Bouloufa}}},
  \bibinfo{author}{\bibfnamefont{O.}~\bibnamefont{{Dulieu}}},
  \bibinfo{author}{\bibfnamefont{D.}~\bibnamefont{{Comparat}}},
  \bibnamefont{and} \bibinfo{author}{\bibfnamefont{P.}~\bibnamefont{{Pillet}}},
  \bibinfo{journal}{\pra} \textbf{\bibinfo{volume}{79}},
  \bibinfo{pages}{021402} (\bibinfo{year}{2009}).

\bibitem[{\citenamefont{Bouloufa et~al.}(2011)\citenamefont{Bouloufa, Pichler,
  Aymar, and Dulieu}}]{2011_PRA_Bouloufa_conversion}
\bibinfo{author}{\bibfnamefont{N.}~\bibnamefont{Bouloufa}},
  \bibinfo{author}{\bibfnamefont{M.}~\bibnamefont{Pichler}},
  \bibinfo{author}{\bibfnamefont{M.}~\bibnamefont{Aymar}}, \bibnamefont{and}
  \bibinfo{author}{\bibfnamefont{O.}~\bibnamefont{Dulieu}},
  \bibinfo{journal}{Phys. Rev. A} \textbf{\bibinfo{volume}{83}},
  \bibinfo{pages}{022503} (\bibinfo{year}{2011}).

\bibitem[{\citenamefont{{Bouloufa} et~al.}(2010)\citenamefont{{Bouloufa},
  {Favilla}, {Viteau}, {Chotia}, {Fioretti}, {Gabbanini}, {Allegrini}, {Aymar},
  {Comparat}, {Dulieu} et~al.}}]{2010MolPh.108.2355B}
\bibinfo{author}{\bibfnamefont{N.}~\bibnamefont{{Bouloufa}}},
  \bibinfo{author}{\bibfnamefont{E.}~\bibnamefont{{Favilla}}},
  \bibinfo{author}{\bibfnamefont{M.}~\bibnamefont{{Viteau}}},
  \bibinfo{author}{\bibfnamefont{A.}~\bibnamefont{{Chotia}}},
  \bibinfo{author}{\bibfnamefont{A.}~\bibnamefont{{Fioretti}}},
  \bibinfo{author}{\bibfnamefont{C.}~\bibnamefont{{Gabbanini}}},
  \bibinfo{author}{\bibfnamefont{M.}~\bibnamefont{{Allegrini}}},
  \bibinfo{author}{\bibfnamefont{M.}~\bibnamefont{{Aymar}}},
  \bibinfo{author}{\bibfnamefont{D.}~\bibnamefont{{Comparat}}},
  \bibinfo{author}{\bibfnamefont{O.}~\bibnamefont{{Dulieu}}},
  \bibnamefont{et~al.}, \bibinfo{journal}{Mol. Phys.}
  \textbf{\bibinfo{volume}{108}}, \bibinfo{pages}{2355} (\bibinfo{year}{2010}),
  \eprint{1005.3283}.

\bibitem[{\citenamefont{Amiot et~al.}(1988)\citenamefont{Amiot, Demtr\"oder,
  and Vidal}}]{1988_JCP_demtroder_Cs2_sepctroscopy}
\bibinfo{author}{\bibfnamefont{C.}~\bibnamefont{Amiot}},
  \bibinfo{author}{\bibfnamefont{W.}~\bibnamefont{Demtr\"oder}},
  \bibnamefont{and} \bibinfo{author}{\bibfnamefont{C.~R.} \bibnamefont{Vidal}},
  \bibinfo{journal}{J. Chem. Phys.} \textbf{\bibinfo{volume}{88}},
  \bibinfo{pages}{5265} (\bibinfo{year}{1988}).

\bibitem[{\citenamefont{{Fioretti} et~al.}(1999)\citenamefont{{Fioretti},
  {Comparat}, {Drag}, {Amiot}, {Dulieu}, {Masnou-Seeuws}, and
  {Pillet}}}]{1999EPJD....5..389F}
\bibinfo{author}{\bibfnamefont{A.}~\bibnamefont{{Fioretti}}},
  \bibinfo{author}{\bibfnamefont{D.}~\bibnamefont{{Comparat}}},
  \bibinfo{author}{\bibfnamefont{C.}~\bibnamefont{{Drag}}},
  \bibinfo{author}{\bibfnamefont{C.}~\bibnamefont{{Amiot}}},
  \bibinfo{author}{\bibfnamefont{O.}~\bibnamefont{{Dulieu}}},
  \bibinfo{author}{\bibfnamefont{F.}~\bibnamefont{{Masnou-Seeuws}}},
  \bibnamefont{and} \bibinfo{author}{\bibfnamefont{P.}~\bibnamefont{{Pillet}}},
  \bibinfo{journal}{Eur. Phys. J. D} \textbf{\bibinfo{volume}{5}},
  \bibinfo{pages}{389} (\bibinfo{year}{1999}).

\bibitem[{\citenamefont{Xie et~al.}(2009)\citenamefont{Xie, Sovkov, Lyyra,
  D.~Li, Bai, Ivanov, Magnier, , and Li}}]{2009_JCP_LiLi_triplet}
\bibinfo{author}{\bibfnamefont{F.}~\bibnamefont{Xie}},
  \bibinfo{author}{\bibfnamefont{V.~B.} \bibnamefont{Sovkov}},
  \bibinfo{author}{\bibfnamefont{A.~M.} \bibnamefont{Lyyra}},
  \bibinfo{author}{\bibfnamefont{S.~I.} \bibnamefont{D.~Li}},
  \bibinfo{author}{\bibfnamefont{J.}~\bibnamefont{Bai}},
  \bibinfo{author}{\bibfnamefont{V.~S.} \bibnamefont{Ivanov}},
  \bibinfo{author}{\bibfnamefont{S.}~\bibnamefont{Magnier}}, ,
  \bibnamefont{and} \bibinfo{author}{\bibfnamefont{L.}~\bibnamefont{Li}},
  \bibinfo{journal}{J. Chem. Phys.} \textbf{\bibinfo{volume}{130}},
  \bibinfo{pages}{051102} (\bibinfo{year}{2009}).

\bibitem[{\citenamefont{{Sofikitis} et~al.}(2009)\citenamefont{{Sofikitis},
  {Horchani}, {Li}, {Pichler}, {Allegrini}, {Fioretti}, {Comparat}, and
  {Pillet}}}]{2009PRA_Sofikitis}
\bibinfo{author}{\bibfnamefont{D.}~\bibnamefont{{Sofikitis}}},
  \bibinfo{author}{\bibfnamefont{R.}~\bibnamefont{{Horchani}}},
  \bibinfo{author}{\bibfnamefont{X.}~\bibnamefont{{Li}}},
  \bibinfo{author}{\bibfnamefont{M.}~\bibnamefont{{Pichler}}},
  \bibinfo{author}{\bibfnamefont{M.}~\bibnamefont{{Allegrini}}},
  \bibinfo{author}{\bibfnamefont{A.}~\bibnamefont{{Fioretti}}},
  \bibinfo{author}{\bibfnamefont{D.}~\bibnamefont{{Comparat}}},
  \bibnamefont{and} \bibinfo{author}{\bibfnamefont{P.}~\bibnamefont{{Pillet}}},
  \bibinfo{journal}{Phys. Rev. A} \textbf{\bibinfo{volume}{80}},
  \bibinfo{pages}{051401} (\bibinfo{year}{2009}).

\bibitem[{\citenamefont{{Lignier} et~al.}(2011)\citenamefont{{Lignier},
  {Fioretti}, {Horchani}, {Drag}, {Bouloufa}, {Allegrini}, {Dulieu}, {Pruvost},
  {Pillet}, and {Comparat}}}]{2011PCCP_Lignier}
\bibinfo{author}{\bibfnamefont{H.}~\bibnamefont{{Lignier}}},
  \bibinfo{author}{\bibfnamefont{A.}~\bibnamefont{{Fioretti}}},
  \bibinfo{author}{\bibfnamefont{R.}~\bibnamefont{{Horchani}}},
  \bibinfo{author}{\bibfnamefont{C.}~\bibnamefont{{Drag}}},
  \bibinfo{author}{\bibfnamefont{N.}~\bibnamefont{{Bouloufa}}},
  \bibinfo{author}{\bibfnamefont{M.}~\bibnamefont{{Allegrini}}},
  \bibinfo{author}{\bibfnamefont{O.}~\bibnamefont{{Dulieu}}},
  \bibinfo{author}{\bibfnamefont{L.}~\bibnamefont{{Pruvost}}},
  \bibinfo{author}{\bibfnamefont{P.}~\bibnamefont{{Pillet}}}, \bibnamefont{and}
  \bibinfo{author}{\bibfnamefont{D.}~\bibnamefont{{Comparat}}},
  \bibinfo{journal}{Phys. Chem. Chem. Phys.} \textbf{\bibinfo{volume}{13}},
  \bibinfo{pages}{18910} (\bibinfo{year}{2011}).

\bibitem[{\citenamefont{Kr\"{a}mer et~al.}(1997)\citenamefont{Kr\"{a}mer, Keil,
  Wang, Bernheim, and Demtr\"{o}der}}]{Kraemer1997391}
\bibinfo{author}{\bibfnamefont{H.-G.} \bibnamefont{Kr\"{a}mer}},
  \bibinfo{author}{\bibfnamefont{M.}~\bibnamefont{Keil}},
  \bibinfo{author}{\bibfnamefont{J.}~\bibnamefont{Wang}},
  \bibinfo{author}{\bibfnamefont{R.}~\bibnamefont{Bernheim}}, \bibnamefont{and}
  \bibinfo{author}{\bibfnamefont{W.}~\bibnamefont{Demtr\"{o}der}},
  \bibinfo{journal}{Chem. Phys. Lett.} \textbf{\bibinfo{volume}{272}},
  \bibinfo{pages}{391 } (\bibinfo{year}{1997}).

\end{thebibliography}
\end{document}